\documentclass[12pt]{article}
\usepackage{pdproc,graphicx} 

\begin{document}

\renewcommand{\thefootnote}{\alph{footnote}}
  
\title{
 SOLAR NEUTRINOS\\
 (with a tribute to John N.\ Bahcall)}

\author{G.L.~FOGLI$^*$, E.~LISI, A.~MARRONE, A.~PALAZZO}
\address{ Dipartimento di Fisica and Sezione INFN,
Via Amendola 173, 70126 Bari, Italy}
 \centerline{\footnotesize $^*$Speaker. \tt gianluigi.fogli@ba.infn.it}

\abstract{John N.\ Bahcall championed solar neutrino physics for many years.
Thanks to his pioneering and long-lasting contributions, this field of research 
has not only reached maturity, but has also opened a new window 
on physics beyond the standard electroweak 
model through the phenomenon of neutrino
flavor oscillations. We briefly outline some recent accomplishments in 
the field, and also discuss a couple of issues that do not 
seem to fit in the ``standard picture,'' namely, the chemical controversy
at the solar surface, and possible implications of recent gallium radioactive
source experiments.}
   
\normalsize\baselineskip=15pt

\begin{figure}[h]
\vspace*{11pt}
\hspace*{0pt}
\begin{center}
\includegraphics[scale=0.8]{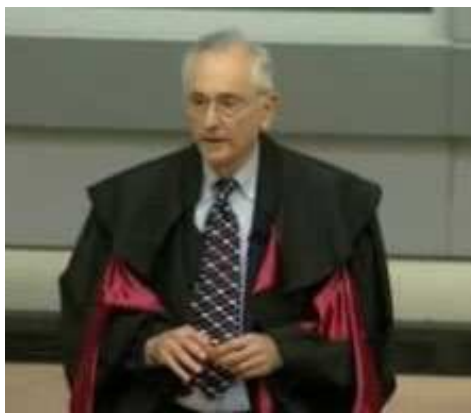}
\end{center}
\caption{John N.\ Bahcall explains ``Why 
the Sun shines'' in his Lectio Magistralis, 
before receiving the Laurea Honoris Causa in Physics
at the University of Milan, Italy (May 6th, 2004).}
\label{fig:1}
\end{figure}

\section{John N.\ Bahcall (1934--2005): Memories of our interactions}

John Bahcall championed solar neutrino physics for many years. His countless
seminars on the solar neutrino problem as a window to new physics are probably
responsible for the interest in neutrino physics of many participants to this 
Workshop (NO-VE 2006) --- and they were definitely so for myself, in the early 1990's. [Milla Baldo Ceolin is also responsible in part... having invited me to give my very first talk on solar neutrinos in the 1993 Workshop on Neutrino Telescopes in Venice.] At that time I was gradually shifting my main interests from electroweak precision physics  to neutrino physics, and
John's classic book on ``Neutrino Astrophysics'' \cite{Book} was a major guidance
in this new (for me) field of research. 

More direct interactions between John and me
started in 1994, first with scientific correspondence on his
famous ``1,000 Standard Solar Models (SSM)'' \cite{1000} (which were used to deal with 
neutrino flux errors and correlations \cite{Corr}), and then
with his acceptance of my former PhD student Eligio Lisi 
as INFN postdoc in his group at the Institute of Advanced Studies in
Princeton.

Interactions with John have always been both scientifically interesting and personally enjoyable. In particular, I remember with great pleasure his participation to our Neutrino Oscillation Workshop 2000 (Otranto,
Italy) \cite{NOW0}, where he managed to come --- and to present preliminary new results from the so-called ``BP 2000'' (Bahcall-Pinsonneault) SSM \cite{BP00} --- despite his many other obligations in that period.

With the same great pleasure I remember his ``Laurea Honoris Causa'' at the
University of Milan, Italy, in May 2004 (see John's picture in Fig.~1), where several participants to this workshop, including myself, were very happy
to be in the Laurea committee celebrating John's outstanding career. That event, as many others, witnessed the close bond of friendship between John and all of us in Italy.

These are just a few facets, from a personal perspective, of his long-life
commitment to science, and to the scientific community. Any of us could add countless examples of such commitment. He was really a leading scientist in our field, and we all miss him greatly. But, his greatest accomplishment remains
with us: solar neutrinos as a window to new physics.

\section{Solar neutrino physics: Established facts}

\begin{figure}[t]
\vspace*{-8pt}
\hspace*{0pt}
\begin{center}
\includegraphics[scale=0.27]{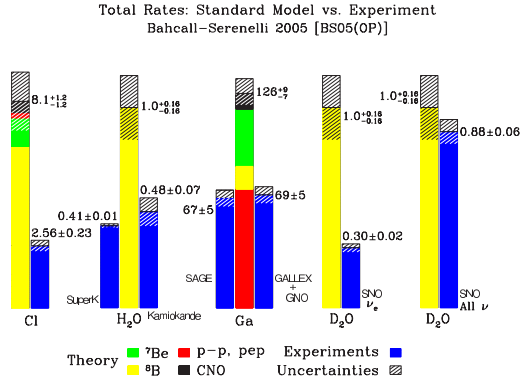}
\end{center}
\vspace*{-14pt}
\caption{Solar neutrino event rates: SSM expectations vs data (2005).
From John's website \protect\cite{JNBW}. }
\label{fig:2}
\end{figure}
  
Four decades (1965-2005) 
of enduring theoretical efforts, difficult experiments, and extraordinary achievements
in solar $\nu$ researches are well summarized by one of John's favorite viewgraphs \cite{JNBW}, reported in Fig.~2. 
The graph shows 
the comparison between SSM expectations (highest bars) and 
data (blue bars) for the observation of the solar neutrino flux in the
Chlorine (Cl) \cite{Home}, Gallium (Ga) \cite{SAGE,GALL,GGNO},
Super-Kamiokande (H$_2$O) \cite{SKso,SK04} and SNO 
(D$_2$O) \cite{SNO1,SNO2,SNOL}, together with their
error bars (shaded areas) and neutrino flux components
(in different colors). All but the rightmost bars show a deficit 
of measured solar $\nu_{e}$ events as compared to no-oscillation
expectations (the famous
solar neutrino problem);  the rightmost bars shows instead no deficit in the
``all $\nu$'' flux from the Sun ($\nu_e+\nu_\mu+\nu_\tau$), implying that
the $\nu_e$ deficit 
 {\em must\/} be due to $\nu_e\to\nu_{\mu,\tau}$ flavor transitions \cite{SNO1,SNO2}.
This phenomenon, theorized long ago by Bruno Pontecorvo \cite{Pont}, requires
massive and mixed neutrinos, and thus appears
to be our first window open beyond the minimal standard model of the electroweak 
interactions. The current solar $\nu$ constraints on this phenomenon 
are briefly summarized below.

\begin{figure}[t]
\vspace*{-4pt}
\hspace*{0pt}
\begin{center}
\includegraphics[width=7.6cm,height=7.3cm]{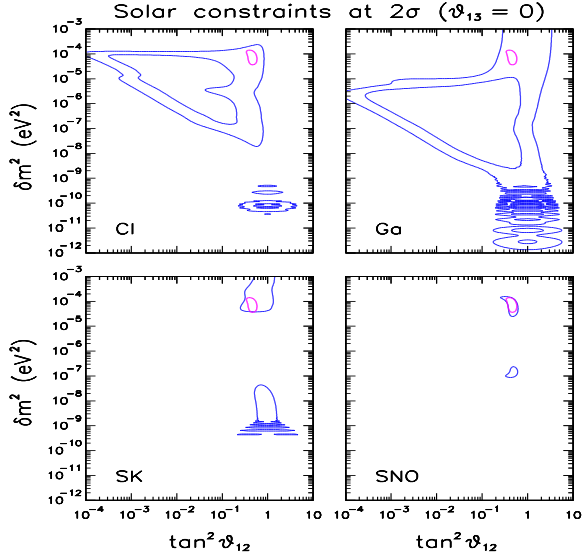}
\end{center}
\vspace*{-2pt}
\caption{Separate (thin) and combined (thick) bounds on the
solar neutrino oscillation parameters from Ga, Cl, SK, and SNO data at
$2\sigma$ (95\% C.L. for 1 d.o.f.), for the case $\theta_{13}=0$ \protect\cite{Revi}. }
\label{fig:3}
\end{figure}
  
\subsection{Status of solar neutrino oscillation parameters}

The dominant flavor transition (``oscillation'') parameters for solar neutrinos
are, in standard notation \cite{PDG4},
the squared mass difference $\delta m^2=m^2_2-m^2_1$ and the mixing angle 
$\theta_{12}$. Subdominant effects can be induced by the mixing angle $\theta_{13}$, which is known to be small \cite{CHOO}. Assuming $\theta_{13}=0$,
the current solar neutrino constraints on the dominant parameters are shown
in Fig.~3 \cite{Revi,Othe}, both from separate experiments (four panels) and in combination
(thicker ``potato'' in each panel).
The combination, universally known as ``large mixing angle'' (LMA)
solution is currently: (1) unique (no multiple solutions); (2) highly
consistent with each data set; and (3) dominated by the SNO experiment,
and to a lesser extent by the SK experiment (which cuts the spurious 
low-$\delta m^2$  solution still allowed by SNO). Lower-energy data 
(Ga and Cl) play instead a role in constraining $\theta_{13}$ (see later).

\begin{figure}[t]
\vspace*{-2pt}
\hspace*{0pt}
\begin{center}
\includegraphics[scale=0.32]{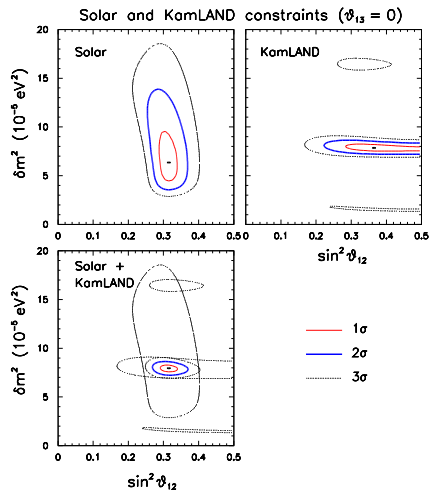}
\end{center}
\vspace*{-8pt}
\caption{Contours of the regions allowed
at 1, 2, and 3$\sigma$ (for 1 d.o.f.) by solar $\nu$ data
(upper left panel), KamLAND data (upper right panel)
and their combination (lower panel) in the case $\theta_{13}=0$.}
\label{fig:4}
\vspace*{-34pt}
\end{figure}

\subsection{Consistency with Standard Solar Model predictions}

As already remarked, the constraints on the
leading parameters $(\delta m^2,\theta_{12})$ are dominated by SNO
and SK data, sensitive to the high-energy ($^8$B) component
of the solar neutrino flux. One can perform 
an analysis of SNO and SK data independent of both the SSM and the 
details of the oscillation phenomenon \cite{Modi}, showing that such data 
constrain the total $^8$B neutrino flux in a range consistent
with SSM predictions \cite{Revi} within $1\sigma$. Moreover, the experimental
error on this flux is a factor of $\sim 2$ smaller than the theoretical SSM one
\cite{Ba05}, implying that the latter does not play anymore a crucial role in
the determination of the LMA parameters. See also \cite{SNOL}.
 
\subsection{Consistency with KamLAND reactor neutrino data}

The KamLAND observation of reactor 
$\bar\nu_e\to\bar\nu_e$ disappearance \cite{Kam1} and associated spectral
distortions \cite{Kam2}, driven by the same leading parameters
$(\delta m^2,\theta_{12})$ as for solar $\nu_e$, has strongly 
increased our confidence in $\nu$ oscillations. Figure~4 \cite{Revi} shows
the very good
consistency between solar and KamLAND constraints on the oscillation
parameters, as well as their complementarity: solar $\nu$ data mainly fix $\sin^2\theta_{12}$ (which is basically measured
by SNO through the charged-to-neutral current event ratio CC/NC \cite{SNOL}),
while KamLAND data mainly fix $\delta m^2$ (through the spectral distortion pattern \cite{Kam2}). Their combination in Fig.~4 can be summarized (with $\pm2\sigma$ errors) as:
\begin{eqnarray}
\delta m^2 &=& 7.92(1\pm0.09)\times 10^{-5} \mathrm{\ eV}^2\ ,\\
\sin^2\theta_{12} &=& 0.314(1^{+0.18}_{-0.15})\ .
\end{eqnarray}
\vspace*{-4pt}These
$\pm 2\sigma$ ranges are not altered for $\theta_{13}\neq 0$
\cite{Revi} (not shown).

\newpage
\subsection{Consistency with expected matter effects}

\begin{figure}[t]
\vspace*{0pt}
\hspace*{0pt}
\begin{center}
\includegraphics[scale=0.52]{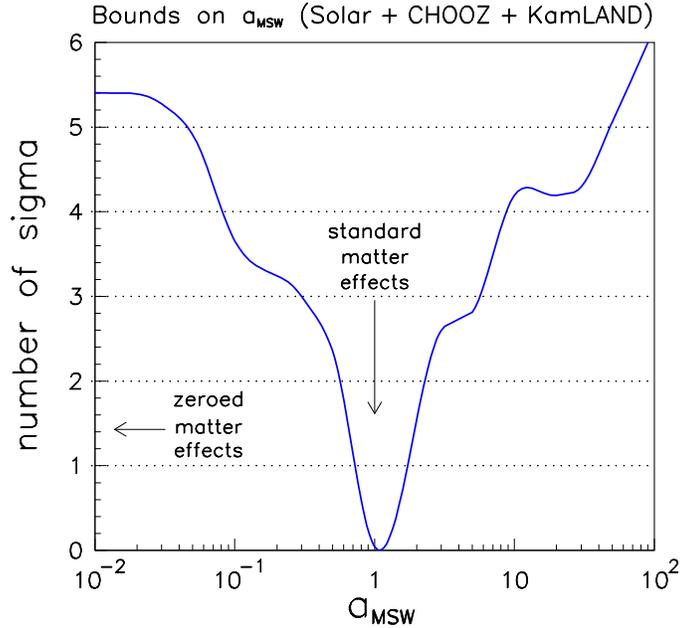}
\end{center}
\caption{Evidence for MSW effects in matter from solar+reactor data \protect\cite{Revi}. See the text for details.}
\vspace*{-20pt}
\label{fig:5}
\end{figure}

After the seminal works by Wolfenstein and by Mikheyev and Smirnov \cite{Matt} (MSW),
it has been realized that even if neutrinos are negligibly absorbed by matter
(a $G^2_F$ effect), their oscillation phases can be significantly modified
by background fermions (a $G_F$ effect). The relevant function governing
the MSW effect in ordinary matter is the so-called neutrino potential $V(x)$,
\begin{equation}
V(x)=\sqrt{2}G_F N_e(x)\ ,
\end{equation}
where $N_e(x)$ is the electron density at the position $x$ along the $\nu$
trajectory. This potential must be definitely taken into account in the
theoretical interpretation of solar $\nu$ data within the
LMA solution (see \cite{MSW2,MSW3} for recent reviews). 

One possible way to test the occurrence of the MSW effect is to
allow a free normalization factor $a_\mathrm{MSW}$ for the potential,
\begin{equation}
V(x) \to a_\mathrm{MSW}V(x)\ ,
\end{equation}
and let the data decide whether $a_\mathrm{MSW}=0$ (no effect) or
$a_\mathrm{MSW}=1$ (standard effect). 

Figure 5 shows the ``number
of sigmas'' of a fit to all solar+reactor data, as a function of $a_\mathrm{MSW}$ \cite{Revi}. The case of no effect is excluded at $>5\sigma$,
while the standard MSW effect is strongly favored, with amplitude constrained
within a factor of $\sim 2$ at $\pm2\sigma$. In a sense, this is an
unconventional measurement
of $G_F$ (although not particularly accurate) through the phenomenon
of neutrino oscillations in matter.

\newpage

\subsection{Synthesis within the three-neutrino mixing framework}
\begin{figure}[t]
\vspace*{0pt}
\hspace*{0pt}
\begin{center}
\includegraphics[scale=0.34]{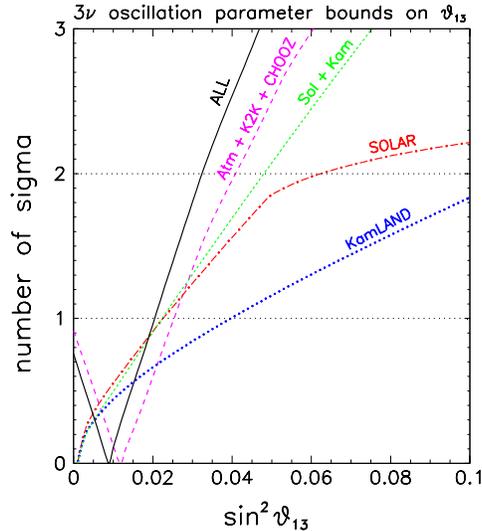}
\end{center}
\vspace*{-5pt}
\caption{Global analysis within the $3\nu$ mixing framework. 
Bounds on $\sin^2\theta_{13}$ from various data sets with increasing
constraining power:
KamLAND, solar, KamLAND+solar, atmospheric+accelerator+CHOOZ, and
all data combined. The overall preference for a nonzero value of $\theta_{13}$
is not statistically significant ($<1\sigma$). From \protect\cite{Revi}.}
\vspace*{-29pt}
\label{fig:6}
\end{figure}

The previous results, although obtained for $\theta_{13}=0$ (effective
$2\nu$ mixing), are not significantly
altered for $\theta_{13}\neq0$ ($3\nu$ mixing), 
due to the nontrivial consistency of
all current $\nu$ oscillation data (both separately and in combination) for
{\em small\/} values of $\theta_{13}$. Indeed, although the current upper
bounds on $\theta_{13}$ are dominated by the null results of the
CHOOZ reactor experiment (in combination
with atmospheric and accelerator data, see e.g., \cite{Revi}), the solar neutrino
constraints are only a factor $\sim2$ weaker. From Fig.~6 one can derive, for instance, the following $2\sigma$ (95\% C.L.) bounds on $\sin^2\theta_{13}$:
\begin{eqnarray}
\sin^2 \theta_{13} &<&0.062 \mathrm{\ (solar)}\ ,\\
                   &<&0.048 \mathrm{\ (solar+KamLAND)}\ ,\\
				   &<&0.032 \mathrm{\ (all\ data)}\ .
\end{eqnarray}
Interestingly, the upper bounds on $\theta_{13}$ from ``solar data only'' 
are due to the interplay between ``low-energy'' (LE) data and 
``high-energy'' (HE) 
data---say, Gallium vs SNO---which measure the solar $\nu_e$
survival probability in different limits \cite{3neu}:
\begin{eqnarray}
P_{ee}^\mathrm{LE}&\simeq & 
(1-2\sin^2\theta_{13})(1-0.5\sin^2 2\theta_{12})\ ,\\
P_{ee}^\mathrm{HE} &\simeq &(1-2\sin^2\theta_{13})\,\sin^2\theta_{12}\ .
\end{eqnarray}
From the above equations it follows that an increase
of $\theta_{13}$ can be compensated by a decrease (increase) of $\theta_{12}$
in low-energy (high-energy) solar neutrino data; then,
as $\theta_{13}$ grows, these diverging shifts in $\theta_{12}$
will eventually become inconsistent with each other and with the data,
leading to the upper bound on $\theta_{13}$ in Eq.~(5).

\newpage\vspace*{-30pt}
\section{Solar neutrino physics: Are there little ``cracks''?}

The previous beautiful and solid facts should not make us overly confident
in our current understanding of solar neutrino physics. Words of caution came 
from John Bahcall himself. At the Neutrino 2002 Conference, when the SSM
predictions for the $^8$B neutrino flux
appeared to be convincingly confirmed by SNO, he stated that 
\begin{quote}
``This is the first time in 40 years of giving talks in solar neutrinos
that it seems to me that the people in the audience are more confident of
the solar model predictions than I am.''
\end{quote}
In addition, one of his favorite quotes (of much more general applicability...)
was:
\begin{quote}
``Half of all three sigma results are wrong.''
\end{quote}
These admonitions suggest an open-minded attitude: one should not be blind to 
small ``cracks'' that might open up in the beautiful solar neutrino construction.

\subsection{Chemical controversy at the solar surface}
One such ``crack'' (the metallicity problem)
had a central role in John's late interests. In a nutshell, it turns
out that, when the newest metal abundances \cite{Aspl} are adopted as SSM input,
the fractional difference between the sound speed profile predicted
by the SSM and the one inferred from  helioseismology
becomes alarmingly large---while it used to be very small
with ``older'' metallicity inputs \cite{Grev}. No clear solution is emerging for this
metallicity problem, as also recognized in John's last research paper \cite{Ba05}. 
Indeed, while previous SSM papers by John always contained a
``recommended'' set of neutrino fluxes and errors, this one \cite{Ba05} leaves an
open choice between (at least) two extreme cases for central values
and errors. 

Luckily, even in the worst case for helioseismology  
(``new'' metal abundances with optimistically small errors), the
estimated $^8$B flux uncertainty in the corresponding SSM \cite{Ba05} is still
a factor $\sim 2$ larger than the experimental one from SNO. Therefore
(see the comments in Sec.~2.2) the solar neutrino bounds on the
LMA parameters 
($\delta m^2,\sin^2\theta_{12}$) remain largely insensitive to the
metallicity problem,
whose disturbing effects seem to be confined
to helioseismology so far. 
However, this problem might surface again in solar neutrino
physics, should the theoretical SSM uncertainties become less
conservative in the future.

\subsection{Issues in gallium radioactive source experiments}

Another small ``discrepancy,'' whose implications
on solar neutrino parameters, to our knowledge,
 have not been discussed before, 
stems from a recent paper \cite{Radi} about the 
radioactive source experiment in SAGE, GALLEX and GNO. These experiments
measure the $\nu$ event rate in Gallium induced by radioactive
$\nu_e$ sources with known intensity, and can thus test the theoretical
cross section for $\nu_e$ absorption in Ga calculated by John Bahcall
in \cite{GaCr}. More precisely, only
the low-energy range of the cross section (say, $<2$ MeV) is tested,
which is relevant for the so-called pp, pep, Be, N, and O contributions
to the solar neutrino flux. The higher energy range ($> 2$ MeV), relevant for 
the B and hep contributions, is instead decoupled from the lower
energy one \cite{Ba02} and is untested by the radioactive source
experiments reported in \cite{Radi}.

Figure~7 shows the various results reported in \cite{Radi}, which can be 
combined  as
\begin{equation}
\mathrm{Ga\ rate\ (radioactive\ source)}: 
\frac{\mathrm{measured}}{\mathrm{predicted}}=0.88\pm0.05
\end{equation}
(shaded band in the figure), where the total error is at $1\sigma$.
In other words, according to the claim in \cite{Radi},
the theoretical Ga cross-section in \cite{GaCr} might be
overestimated by a factor 1/($0.88\pm0.05$) (at least at low energy):
not a negligible effect ($>2\sigma$).

It is then tempting to see what happens if the estimated Ga cross
section in \cite{GaCr} is ``renormalized'' ad hoc by a factor $0.88\pm 0.05$ 
for energies below $\sim2$~MeV. 
We can expect, from the comments
to Eqs.~(8) and (9), that a significant change in the Ga predictions will 
at least alter the bounds on $\theta_{13}$ derived
from solar neutrino data [Eq.~(5)].
In the following we revisit such bounds, by comparing the two cases
with ``standard Ga
cross section'' and with ad hoc ``renormalized Ga cross section''
 [$\sigma_\mathrm{Ga}\to\sigma_\mathrm{Ga}\times (0.88\pm 0.05)$ for
 $E_\nu<2$~MeV].
Our results should be intended as a preliminary exploration of this
issue.

\begin{figure}[t]
\vspace*{-30pt}
\hspace*{0pt}
\begin{center}
\includegraphics[scale=0.45]{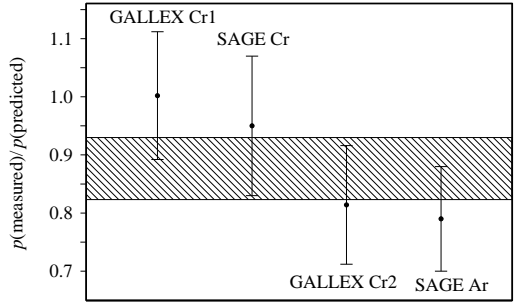}
\end{center}
\caption{Ratio of measured to predicted event rates from radioactive
source experiments in Ga detectors. The shaded band correspond to the
combined result $0.88\pm0.05$. From \protect\cite{Radi}.}
\label{fig:7}
\end{figure}

 Figure~8 shows two key quantities as a function of $\sin^2\theta_{12}$:
the Ga event rate in solar $\nu$ experiments, normalized to SSM \cite{BSB5,Revi}
expectations, and the CC/NC event ratio in SNO. These quantities,
representative of low- and high-energy $\nu$ observables, 
are plotted both for $\sin^2\theta_{13}=0$ (solid) and $\sin^2\theta_{13}=0.05$
(dashed). The horizontal bands represent the corresponding experimental
data at $\pm 1\sigma$. The two panels differ only for the Ga cross
section: standard (left) and ``renormalized'' at low energy (right);
therefore, only the Ga observables change from left to right (not CC/NC).

\begin{figure}[t]
\vspace*{0pt}
\hspace*{0pt}
\begin{center}
\includegraphics[scale=0.64]{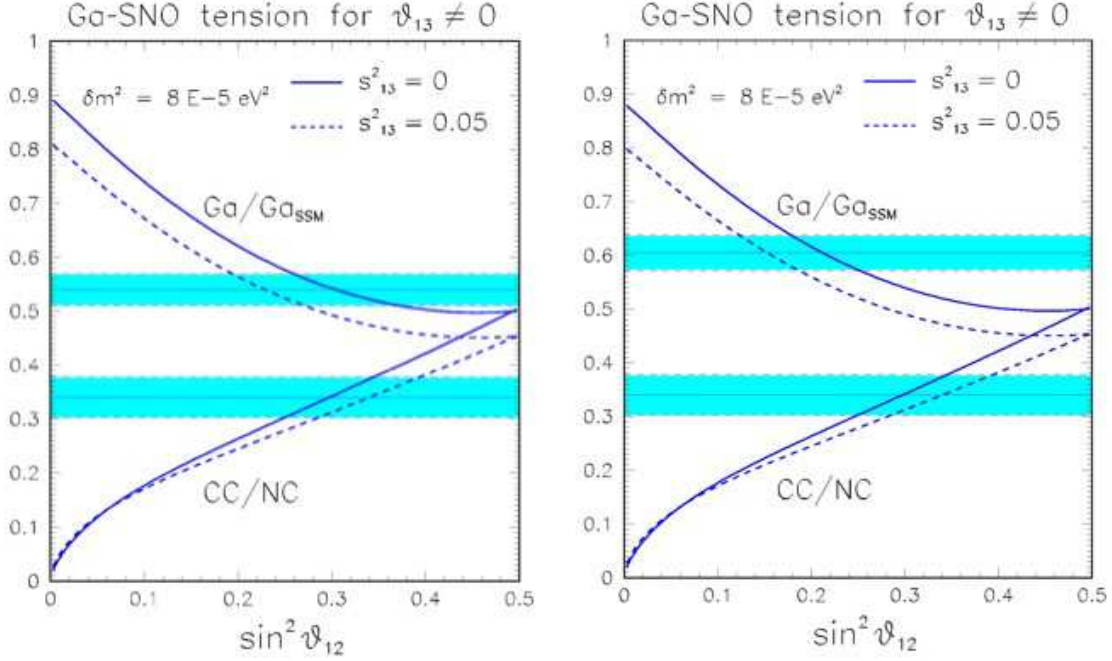}
\end{center}
\vspace*{-4pt}
\caption{Comparison of data and prediction for Gallium and SNO observables,
for standard and ``renormalized'' Ga cross section (left and right plots, respectively). See the text for details.}
\label{fig:8}
\end{figure}

In the left plot of Fig.~8 one can see
 that, for $\sin^2 \theta_{12}\simeq 0.3$, both solid curves
 agree very well with the data, i.e., there
is high consistency between Ga and SNO information at
$\sin^2\theta_{13}=0$. For $\sin^2\theta_{13}=0.05$, however, the
(dashed) curves cross the data bands at different values of $\sin^2\theta_{12}$ (about 0.23 for Ga rate and 0.34 for CC/NC).
This mismatch leads to an upper bound on $\sin^2\theta_{13}$ as in 
Eq.~(5).

In the right plot of Fig.~8 (``renormalized'' cross section),
this mismatch exists already at $\sin^2\theta_{13}=0$, and becomes
worse for $\sin^2\theta_{13}>0$: there is no value
of $\sin^2\theta_{12}$ which accommodates well both Ga and SNO data
with their predictions. In this case, there is a disturbing
 ``tension'' between low- and high-energy solar $\nu$ data. The tension
 would be even stronger if the new (lower) solar metallicity \cite{Ba05}
 were adopted, since it would further decrease the expected
 Ga rate (not shown).

Figure~9 shows this tension in an alternative way, by comparing
the regions allowed at $1\sigma$ by Gallium data (slanted band)
and SNO data (closed region) in the usual mass-mixing plane, for
four increasing values of $\sin^2\theta_{13}$. As in Fig.~8, the
left (right) panels refer to standard (``renormalized'') Gallium cross section.
For standard cross section (left), the Ga and SNO allowed regions fully
overlap for $\sin^2\theta_{13}=0$, but they increasingly separate for
increasing values of $\sin^2\theta_{13}$. This tension leads
to meaningful upper bounds on $\sin^2\theta_{13}$ [Eq.~(5)]. 
For ``renormalized'' cross section, however, there is no 
overlap between Ga and SNO regions at $1\sigma$, even at $\sin^2\theta_{13}=0$:
there is always ``tension''.
Therefore, at face value, one would obtain a formally stronger upper
bound on $\sin^2\theta_{13}$ from solar data (not shown), at the price of
a worse best-fit at $\sin^2\theta_{13}=0$. 

In conclusion, if the claim in favor of a lower Ga cross section \cite{Radi}
is valid, the current good agreement between low and high energy solar
neutrino data would be somewhat spoiled, and the solar $\nu$ indication
for small $\theta_{13}$ would become stronger at face value,
but also more ambiguous.
Therefore, we think that further work is required to interpret
the results in \cite{Radi} before including them in global analyses (through, e.g., a
reduced Ga cross section for $E_\nu<2$ MeV).
Certainly, such results provide one more motivation to explore
the low-energy spectrum of solar neutrinos and to revisit the $\nu_e$
absorption cross section in gallium: two topics 
(among many others) where John's contribution will be greatly missed.

\begin{figure}[t]
\vspace*{13pt}
\hspace*{0pt}
\begin{center}
\includegraphics[scale=0.67]{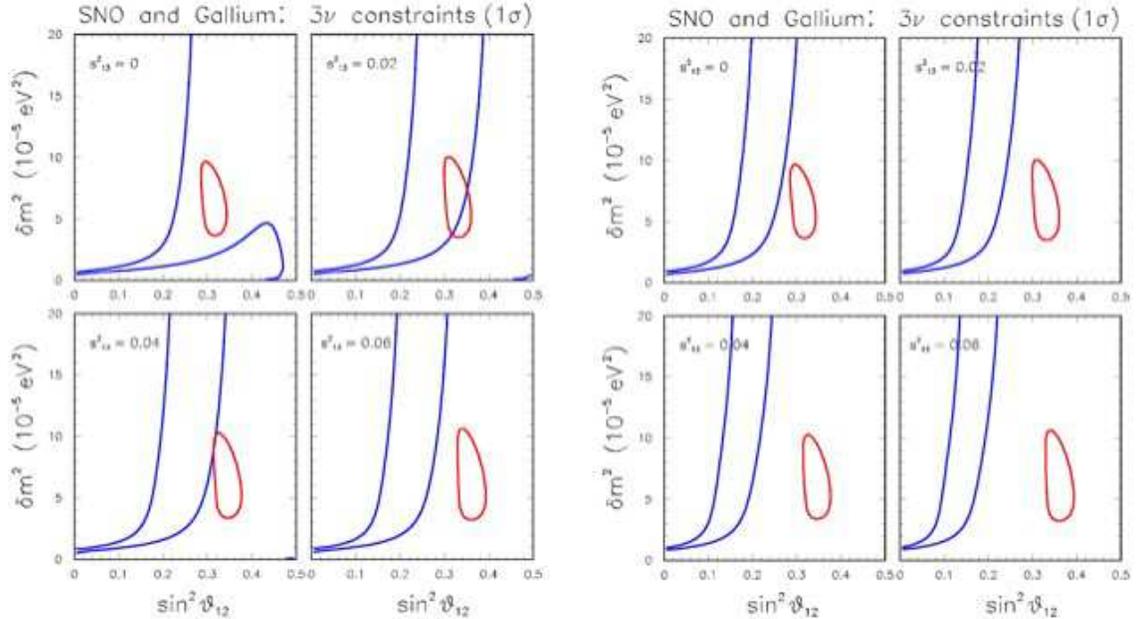}
\end{center}
\vspace*{-14pt}
\caption{Regions separately allowed at $1\sigma$ in the mass-mixing plane by SNO
(closed regions) and Gallium data (bands), for 
$\sin^2\theta_{13}=0,\, 2,\, 4,\, 6\times 10^{-2}$.
The four plots on the left (right) refer to standard (``renormalized'')
Ga cross section. See the text for details.  }
\label{fig:9}
\end{figure}

\section{Conclusions}

We have reviewed the status and the successes of solar neutrino research,
in the light of the outstanding contributions by John Bahcall. Solar neutrinos
provide us with solid evidence in favor of $\nu$ masses and mixing,
in a beautiful synthesis of physics and astrophysics. But, as for any synthesis, we are eager to go beyond it: more accurate studies might be already
revealing something unexpected, e.g., concerning
the metallicity problem, or the Gallium cross section. Time will tell
us if these or other ``disturbances''
 will disappear as
accidental fluctuations, or will instead
evolve as new, independent ``solar neutrino problems''
requiring new (astro)physics.

\newpage
\section{Acknowledgements}
  G.L.F.\ and E.L.\ thank Milla Baldo Ceolin for kind hospitality at
  the NO-VE 2006 Workshop in  Venice. This work is supported in part by the Italian INFN and MIUR through the ``Astroparticle Physics'' research project.


\begin{thebibliography}{99}
  \bibitem{Book} J.N.\ Bahcall, {\em Neutrino Astrophysics\/} 
  (Cambridge University Press, Cambridge, England, 1989).
  
  \bibitem{1000} J.N.\ Bahcall and R.K.\ Ulrich, 
  Rev.\ Mod.\ Phys.\ {\bf 60}, 297 (1988).
  
  \bibitem{Corr} G.L.~Fogli and E.~Lisi,
  Astropart.\ Phys.\  {\bf 3}, 185 (1995).
  
  \bibitem{NOW0} Neutrino Oscillation Workshop (NOW) 2000 (Otranto,
  Italy, 2000). Website: www.ba.infn.it/$^\sim$now2000. Proceedings:
  Nucl.\ Phys.\ B (Proc.\ Suppl.) {\bf 100} (2001), 401 pp., edited by
  G.L.\ Fogli.
  
  \bibitem{BP00} J.N.~Bahcall, M.H.~Pinsonneault and S.~Basu, 
   Astrophys.\ J.\  {\bf 555}, 990 (2001).

  \bibitem{JNBW} J.N.~Bahcall website, www.sns.ias.edu/$^\sim$jnb
  (Solar Neutrinos, Viewgraphs).

\bibitem{Home}  Homestake Collaboration,
                B.T.~Cleveland, T.~Daily, R.~Davis Jr., J.R.~Distel,
                K.~Lande, C.K.~Lee, P.S.~Wildenhain, and
                J.~Ullman, Astrophys.\ J.\  {\bf 496}, 505 (1998).
                
\bibitem{SAGE}  SAGE Collaboration,
                J.N.~Abdurashitov {\it et al.},
                J.\ Exp.\ Theor.\ Phys.\  {\bf 95}, 181 (2002)
                [Zh.\ Eksp.\ Teor.\ Fiz.\  {\bf 95}, 211 (2002)].

\bibitem{GALL}  GALLEX Collaboration, W.~Hampel {\it et al.},
                Phys.\ Lett.\ B {\bf 447}, 127 (1999).

\bibitem{GGNO}  Gallium Neutrino Observatory (GNO) Collaboration,
				M.\ Altmann {\em et al.}, 
				Phys.\ Lett.\ B {\bf 616}, 174 (2005).


\bibitem{SKso}  SK Collaboration, S.~Fukuda {\it et al.},
                Phys.\ Rev.\ Lett.\  {\bf 86}, 5651 (2001);
                Phys.\ Rev.\ Lett.\  {\bf 86}, 5656 (2001);
                Phys.\ Lett.\ B {\bf 539}, 179 (2002).

\bibitem{SK04}  SK Collaboration, 
                M.B.~Smy {\it et al.},
                Phys.\ Rev.\ D {\bf 69}, 011104 (2004).

\bibitem{SNO1}  SNO Collaboration,
                Q.R.~Ahmad {\it et al.},
                Phys.\ Rev.\ Lett.\  {\bf 87}, 071301 (2001);
                Phys.\ Rev.\ Lett.\  {\bf 89}, 011301 (2002);
                Phys.\ Rev.\ Lett.\  {\bf 89}, 011302 (2002).
                
\bibitem{SNO2}  SNO Collaboration, S.N.\ Ahmed {\em et al.},
                Phys.\ Rev.\ Lett.\  {\bf 92}, 181301 (2004).

\bibitem{SNOL}	SNO Collaboration, B.~Aharmim {\it et al.},  
				Phys.\ Rev.\ C {\bf 72}, 055502 (2005).
  
\bibitem{Pont}  B.~Pontecorvo, Zh.\ Eksp.\ Teor.\ Fiz.\ {\bf
                53}, 1717 (1968) [Sov.\ Phys.\ JETP {\bf 26}, 984 (1968)].

\bibitem{PDG4}  Review of Particle Physics,
                S. Eidelman {\em et al.}, Phys. Lett. B {\bf 592}, 1 (2004).

\bibitem{CHOO}  CHOOZ Collaboration, M.~Apollonio {\it et al.},
                Phys.\ Lett.\ B {\bf 466}, 415 (1999);
                Eur.\ Phys.\ J.\ C {\bf 27}, 331 (2003).

  \bibitem{Revi}  G.L.~Fogli, E.~Lisi, A.~Marrone and A.~Palazzo,
  hep-ph/0506083, invited Review to appear in Prog.\ Part.\ Nucl.\ Phys. (2006).
  
  \bibitem{Othe}	 Recent global analyses of 
  solar neutrino data have also appeared,
  e.g., in \cite{SK04,SNOL} and in:
    J.N.~Bahcall, M.C.~Gonzalez-Garcia and C.~Pena-Garay,
  JHEP {\bf 0408}, 016 (2004);
  M.~Maltoni, T.~Schwetz, M.A.~Tortola and J.W.F.~Valle,
  New J.\ Phys.\  {\bf 6}, 122 (2004);
   P.C.~de Holanda and A.Yu.~Smirnov,
  Astropart.\ Phys.\  {\bf 21}, 287 (2004);
  A.~Bandyopadhyay, S.~Choubey, S.~Goswami, S.T.~Petcov and D.P.~Roy,  
  Phys.\ Lett.\ B {\bf 608}, 115 (2005);
  A.~Strumia and F.~Vissani,  Nucl.\ Phys.\ B {\bf 726}, 294 (2005).
  
  \bibitem{Modi}  F.L.~Villante, G.~Fiorentini and E.~Lisi,  
  Phys.\ Rev.\ D {\bf 59}, 013006 (1999).
  
  \bibitem{Ba05}  J.N.~Bahcall, A.M.~Serenelli and S.~Basu,
  astro-ph/0511337.
  
\bibitem{Kam1}	KamLAND Collaboration, K.~Eguchi {\it et al.},
			  	Phys.\ Rev.\ Lett.\  {\bf 90}, 021802 (2003).

\bibitem{Kam2}	KamLAND Collaboration, T.\ Araki {\it et al.},
			  	Phys.\ Rev.\ Lett.\  {\bf 94}, 081801 (2005).	
  
\bibitem{Matt}  L.~Wolfenstein,
                Phys.\ Rev.\ D {\bf 17}, 2369 (1978);
                S.~P.~Mikheev and A.~Yu.\ Smirnov,
                Yad.\ Fiz.\ {\bf 42}, 1441 (1985)
                [Sov.\ J.\ Nucl.\ Phys.\ {\bf 42}, 913 (1985)].

\bibitem{MSW2}	A.Yu.~Smirnov,
				Phys.\ Scripta {\bf T121}, 57 (2005)

\bibitem{MSW3}	G.~Fogli and E.~Lisi,  
				New J.\ Phys.\  {\bf 6}, 139 (2004).

  \bibitem{3neu} S.~Goswami and A.Yu.~Smirnov,
Phys.\ Rev.\ D {\bf 72}, 053011 (2005).
  
  \bibitem{Aspl} M.~Asplund, N.~Grevesse, A.J.~Sauval, C.~Allende Prieto and D.~Kiselman,
Astron.\ Astrophys.\  {\bf 417}, 751 (2004).
  
  \bibitem{Grev} N.~Grevesse and A.J.~Sauval,
Space Sci.\ Rev.\  {\bf 85}, 161 (1998).  
  
  \bibitem{Radi} SAGE Collaboration, J.N.~Abdurashitov {\it et al.},
  Phys.\ Rev.\ C {\bf 73}, 045805 (2006).
  
  \bibitem{GaCr} J.N.~Bahcall,
Phys.\ Rev.\ C {\bf 56}, 3391 (1997).
  
  \bibitem{Ba02} J.N.~Bahcall, M.C.~Gonzalez-Garcia and C.~Pena-Garay,
  Phys.\ Rev.\ C {\bf 66}, 035802 (2002).
  
  \bibitem{BSB5} J.N.~Bahcall, A~M.~Serenelli and S.~Basu,
		Astrophys.\ J.\  {\bf 621}, L85 (2005).
  
  
  
\end{thebibliography}
\end{document}